\documentclass[sigconf]{acmart}

\usepackage[ruled,vlined,linesnumbered]{algorithm2e}

\usepackage{algorithmic}
\usepackage{amsmath}

\usepackage{amsthm}
\usepackage{multirow}

\usepackage{graphicx}
\graphicspath{ {./images/} }

\usepackage{balance}

\AtBeginDocument{%
  }

\copyrightyear{2024}
\acmYear{2024}
\setcopyright{acmlicensed}\acmConference[CIKM '24]{Proceedings of the 33rd ACM International Conference on Information and Knowledge Management}{October 21--25, 2024}{Boise, ID, USA}
\acmBooktitle{Proceedings of the 33rd ACM International Conference on Information and Knowledge Management (CIKM '24), October 21--25, 2024, Boise, ID, USA}
\acmDOI{10.1145/3627673.3680059}
\acmISBN{979-8-4007-0436-9/24/10}

\begin{document}

\title{DivNet: Diversity-Aware Self-Correcting Sequential Recommendation Networks}


\author{Shuai Xiao}
\affiliation{%
  \institution{Alibaba Group}
  \city{Shanghai}
  \country{China}}
\email{shuai.xsh@alibaba-inc.com}

\author{Zaifan Jiang}
\affiliation{%
  \institution{Alibaba Group}
  \city{Beijing}
  \country{China}}
\email{zaifan.jzf@alibaba-inc.com}

\renewcommand{\shortauthors}{Shuai Xiao and Zaifan Jiang}

\begin{abstract}
As the last stage of a typical \textit{recommendation system}, \textit{collective recommendation} aims to give the final touches to the recommended items and their layout so as to optimize overall objectives such as diversity and whole-page relevance. In practice, however, the interaction dynamics among the recommended items, their visual appearances and meta-data such as specifications are often too complex to be captured by experts' heuristics or simple models. To address this issue, we propose a \textit{\underline{div}ersity-aware self-correcting sequential recommendation \underline{net}works} (\textit{DivNet}) that is able to estimate utility by capturing the complex interactions among sequential items and diversify recommendations simultaneously. Experiments on both offline and online settings demonstrate that \textit{DivNet} can achieve better results compared to baselines with or without collective recommendations. 
\end{abstract}



\maketitle

\section{Introduction}
Modern \textit{recommendation systems} are complex, often consisting of multiple stages including matching (a.k.a. retrieval), coarse-to-fine-grain tiered ranking, and \textit{collective recommendation} systems. Given user queries (e.g., a search query or a user session), the matching module quickly identifies a moderate-sized set of relevant candidates from a vast amount (millions or billions) of items in the inventory. Tiered ranking modules then estimate item-wise scores, such as CTR and CVR, based on which items will be ranked w.r.t. their relevancy to the target user. Before surfacing the recommendations to the user, a collective recommendation module is usually employed to optimize the value delivery. Without this module, the recommendation systems may suffer from severe problems, such as the discrepancy between training and inference, lack of diversity and so on. The discrepancy between training and inference means the model loss considers interactions between ranked items when training while the item is scored point-wisely or the context for this item has changed during inference. This discrepancy can lead to unreliable prediction performance. Given certain queries, similar items usually tend to have high scores, resulting in over-concentration of recommended items.
Recently, collective recommendation module~\cite{steck2018calibrated,nassif2018diversifying,teo2016adaptive,wilhelm2018practical} is introduced into recommendation systems to further optimize the item layout, which can boost the performances of recommendation system, such as improving user satisfactions and click conversion rates. Collective recommendations try to improve the whole-page attraction by modeling the item interaction dynamics.

Collective recommendation poses unique technical challenges.
\noindent
\renewcommand{\theenumi}{\Roman{enumi}}
\begin{enumerate}
\item First, complex interactions exist between items in different positions. Similar items shown in top positions may decrease or increase the click probability of subsequent items (which we call a self-exciting process between items). These depend on subtle causalities that it's hard to define explicitly using rules beforehand. For instance, some users may prefer a diverse recommendation set while a concentrated one is preferred for other users. Besides, distinct combinations of items result in different user experiences. Items combinations with compatible colors or text descriptions can grasp users' attention which in turn improves the click-through rate. Items with a short distance have stronger interactions than those with a long one. 
\item Secondly, sequence exposure bias also exists as only a small fraction of possible item combinations have been shown to users, which leads to the estimating of out-of-distribution item combinations unreliable. 
\end{enumerate}

Recent literature~\cite{steck2018calibrated} addresses this issue by introducing Bayesian prior distribution of users' preferences into the optimizing objective. The resulting recommendation sets are close to prior distributions of users' preferences. However, it requires enough data accumulation to estimate an accurate prior distribution which is unsuitable for scenarios, like cold-start recommendations and short-term advertising or marketing activities. Nassif et al.~\cite{nassif2018diversifying,teo2016adaptive} try to balance the click-through rate and item diversity by designing a sub-modular function. Similarly, Wilhelm et al.~\cite{wilhelm2018practical} employ determinantal point processes to diversify YouTube recommendation by computing pair-wise item similarity, which is inefficient to capture high-order item interactions and is computationally expensive. All the above methods assume the element-wise utility of items remain unchanged regardless of the contextual items.   These methods are inflexible to model complex interactions within items, between items and users, items and positions.
The PRM~\cite{pei2019personalized} employs a transformer structure to learn the global interactions from all the combinations of item pairs on the list. Seq2Slate~\cite{bello2018seq2slate} employs pointer-net~\cite{vinyals2015pointer} (a sequence-to-sequence network) to capture item interactions where a RNN encoder embeds all items and a RNN decoder iteratively select items using attention mechanism. The PRM and Seq2Slate is not diversity-aware and doesn't model diversity explicitly.

To address the above concerns, we propose a \textit{self-correcting sequential recommendation networks} (\textit{DivNet}) which can estimate the item utility sequentially and diversify recommendation set simultaneously. The \textit{DivNet} firstly projects items and user features with contextual items into a hidden space using self-attention networks. Then \textit{DivNet} sequentially selects items by considering influences of previously-selected items and passes its influence to subsequent items. During this selecting process, the \textit{DivNet} estimates the item utility and computes the similarity between candidates and existing items. The self-correcting module aims at maximizing total utility and improve diversity among selected items. To reduce the effect of exposure bias, we let the \textit{DivNet} search in the nearby region of original exposed sequences by adding a supervised regularization.

We conduct enormous ablation tests in industrial applications. Results of offline evaluation and deployment on commercial platforms demonstrate its superior performances compared to baselines with/without collective recommendations. What's more, the proposed method is practical as it can support large-scale queries in real-world recommendation systems.

\section{Related Works}
\subsection{Recommendation System}
\begin{figure}[htb]
\centering
\includegraphics[width=.35\textwidth]{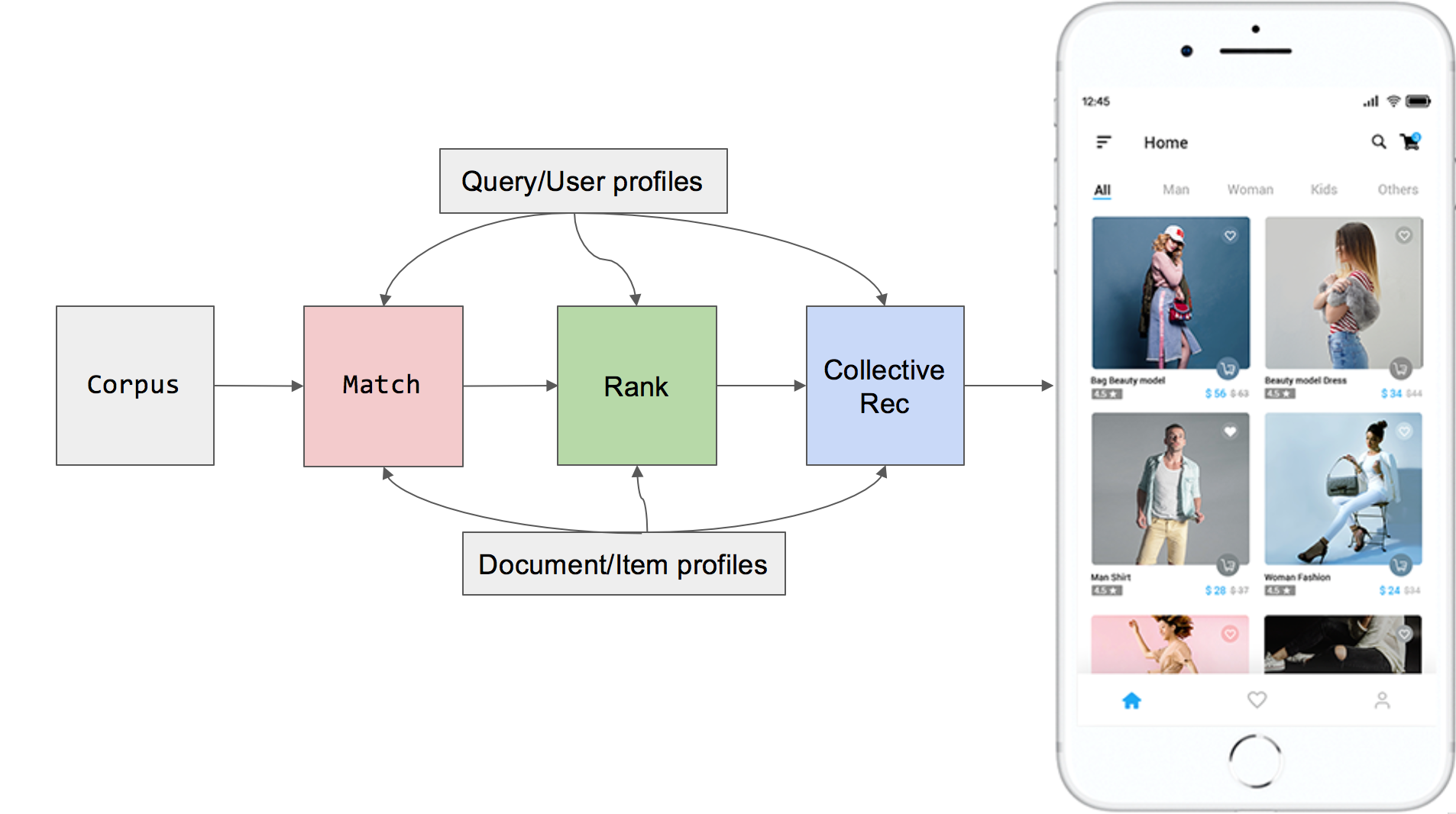}
\caption{Typical architecture of industrial-level recommendation systems. It includes feature layer, matching, ranking. The last layer in the yellow box is collective recommendation which puts the final touches to the recommended items and their layout so as to optimize objectives such as click-through-rate or dwell time.}
\label{fig:resys}
\end{figure}

The architecture of industrial-level recommendation systems usually have matching, ranking (coarse and fine ranking) modules as shown in Figure~\ref{fig:resys}.
The difference between matching~\cite{covington2016deep} and ranking is that matching mainly emphasizes on inference speed with proper precision while ranking lays stress on ranking precision.

Traditional approaches of learning-to-rank estimate the utility of each item, \emph{either} individually \cite{agarwal2019general} \emph{or} as a pair \cite{Rendle09} or $k$-tuple \cite{Yang2011}, by learning a ranking function. Items are then ranked in a descending order according to the estimated utility, and the top-$k$ list is presented to the user. 
These methods solve the recommendation approximately, and they often work but only to the extent where (1) items can be considered independent such that the utility of the bundle \emph{as a whole} are decomposable as the sum (or some linear function) of the individual utilities; and (2) the attraction of the whole page depends entirely on the content of these $k$ items, regardless of their layout. In any realistic scenario, these two assumptions are regarded too restrictive and cannot reflect the truthful reality ~\cite{ding2019whole,wang2016beyond,zhao2018deep,wang2017efficient}. For example, numerous user studies have found that there exist complex interaction patterns not only between items but also among the content of the items, their visual appearances and user's attention distribution over the screen \cite{JooWon12}.

\subsection{Collective Recommendation}
Collective recommendation is used to address drawbacks discussed above and output optimal item layout. the LambdaMART~\cite{burges2010ranknet} incorporate the whole list of items and try to find an optimal list through the listwise ranking loss. Nassif et al.~\cite{nassif2018diversifying,teo2016adaptive} try to balance the click-through rate and item diversity by designing a sub-modular function.  
The sub-modular function guarantees that the function gain decreases as the number of selected items in the same category increases gradually
The items can be selected greedily by maximizing the sub-modular objective. Similarly, Wilhelm et al.~\cite{wilhelm2018practical} employ determinantal point processes to diversify YouTube recommendation by computing pair-wise item similarity. This approach is inefficient in capturing high-order item interactions and is computationally expensive. All the above methods assume the utility of single items remains constant regardless of surrounding items. The estimation of item utility and collective ranking is divided into two stages. Actually, the utility changes when similar items have appeared previously. Our work estimates item utility considering the contextual items and diversify recommendations at the same time. The PRM~\cite{pei2019personalized} employs a transformer structure to learn the global interactions from all the combinations of item pairs on the list. Seq2Slate~\cite{bello2018seq2slate} employs pointer-net~\cite{vinyals2015pointer} (a sequence-to-sequence network)to capture item interactions where a RNN encoder embeds all items and a RNN decoder iteratively select items using attention mechanism. Seq2Slate is not diversity-aware and doesn't model diversity explicitly.

\subsection{Self-correcting Process}
Self-correcting process \cite{ogata1998space,hawkes1971spectra,aalen2008survival} is a statistical framework which describes the probability of events happening in the temporal and/or spatial fields. For recommendation system, the propensity that users are satisfied with recommended items constitutes the intensity in self-correcting process. The item interactions when shown sequentially in positions of the display page gives a well-defined spatial self-correcting process. 
The influence can be positive or negative, depending on item interactions and user preferences.

\section{Collective Recommendation via DivNet}
In this section, we'll give details of network structures of \textit{DivNet} whose inputs are ranked items $\mathbf{A} = \{a_i\}_{i=1}^N$ from upstream ranking modules and sequentially outputs item layout $\pi_{\mathbf{A}}$, a permutation of $\mathbf{A}$ and the motivations behind these network structures.

\subsection{Model Architecture}
\begin{figure*}[htb]
\centering
\includegraphics[width=0.6\textwidth]{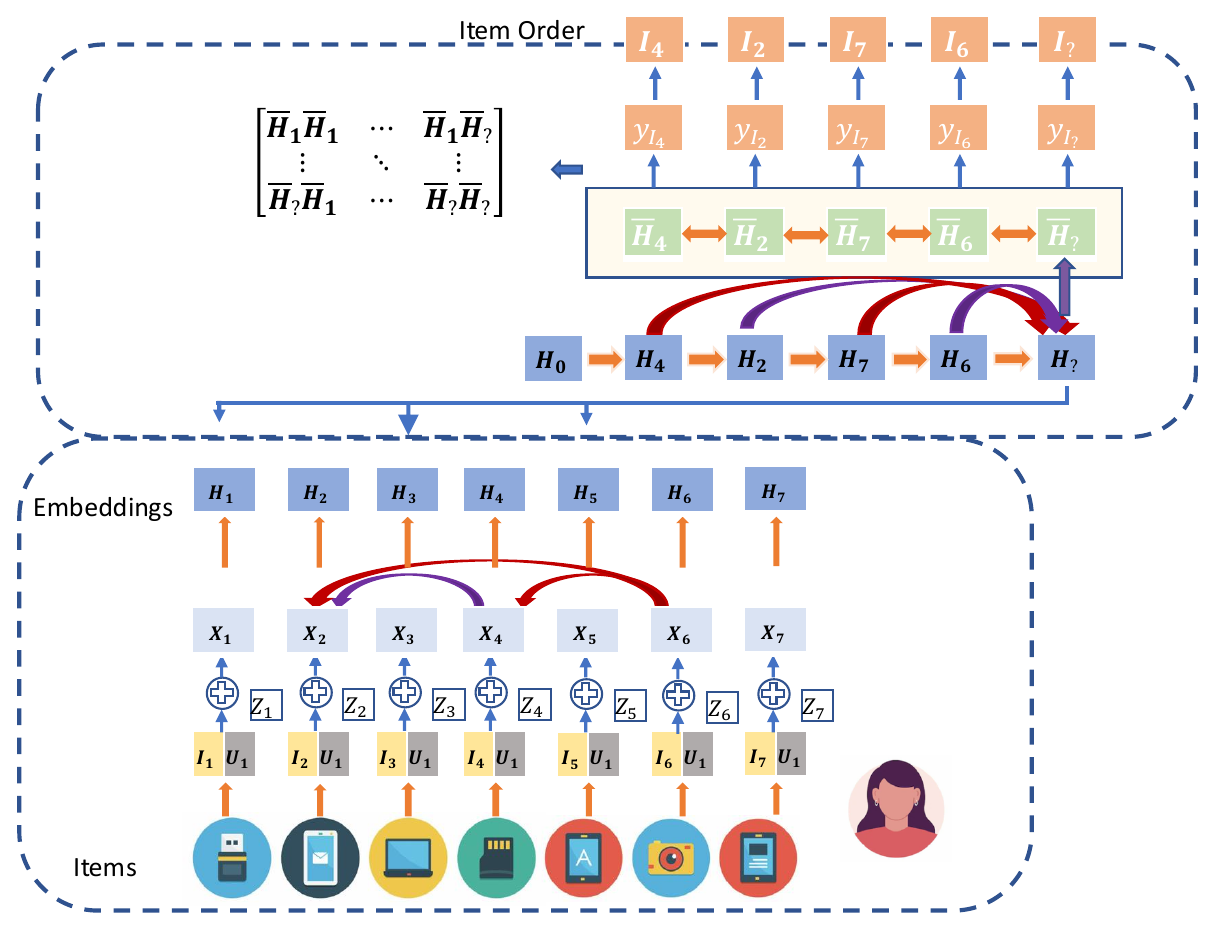}
\caption{The network structures of the proposed self-correcting sequential recommendation network $\textit{DivNet}$. It first projects items into a hidden space with awareness of the existence of contextual items. Then \textit{DivNet} sequentially selects items by considering influences of previously-selected items and passes their influence to subsequent items. Note that \textit{DivNet} considers the sequential self-correcting of item interaction explicitly as shown in the decoding network when recommendation items are sequentially viewed by users.}
\label{fig:DivNet}
\end{figure*}

The \textit{DivNet} follows a sequence-to-sequence paradigm as shown in Figure~\ref{fig:DivNet}. The ranked items from upstream are embedded in a hidden low-dimensional space with information from all other items collected in a message-passing way. Therefore, individual item embedding contains information about the accompanying items with whom they cooperate or compete. 

\textbf{Input Transformation}
In order to let the model be aware of the spatial information of page layout, the positions can be encoded in a deterministic way by the following equations
\begin{align}\label{equ:pos_embedding}
    \mathbf{Z}_j^i= \begin{cases}\sin\big(\frac{i}{10000^{2*(j//2)/L}}\big), &j \in Even \quad Number  \\
                    \cos\big(\frac{i}{10000^{2*(j//2)/L}}\big), &j \in Odd\quad Number
        \end{cases}
\end{align}
where $\mathbf{Z}_j^i$ is the $j$-th value of position $i$ embedding. $\mathbf{Z}\in \mathcal{R}^{|A|\times L}$ is the position embedding matrix where $|\mathbf{A}|$ is the total number of positions and $L$ is the dimension of position embedding.

Besides spatial embedding, the item features of set $\mathbf{A}$ are represented as feature matrix $\mathbf{I}\in \mathcal{R}^{|\mathbf{A}|\times M}$ where  $M$ is the dimension of item embedding. The $i$-th row of $\mathbf{I}$ contains $i$-th item's continuous features and trainable embedding of discrete features. To encode query (user context in recommendation) for personalized recommendation, features of user context denoted by $\mathbf{U}\in\mathcal{R}^{1\times K}$, are concatenated to item features. Then the concatenation is added with spatial embedding.
Therefore, the features of the input can be expressed as:
\begin{align}
    \mathbf{X} &= [\mathbf{I}, \mathbf{U}] + \mathbf{Z}
\end{align}
where $[\cdot]$ is the concatenation operator along the 2ed dimension with broadcasting which means that query embedding $\mathbf{U}$ is repeated $|\mathbf{A}|$ times along the 1st dimension. $\mathbf{X}\in \mathcal{R}^{|\mathbf{A}|\times (M+K)}$ is an intermediate variable where each row corresponds to the embedding of one item in the item sequence $\mathbf{A}$.

After initial feature concatenation and addition of spatial feature, the items $\mathbf{A}$ are projected into a hidden feature space with awareness of contextual items.  The calculation of projection follows a self-attention mechanism governed by the following equation:
\begin{align}\label{equ:self-attention-softmax}
    \mathbf{E_e} &= \frac{\text{Softmax}\big(\mathbf{Q_e} * \mathbf{K_e}^T)}{\mathbf{\sqrt{D_K}}} \mathbf{V_e} \\\label{equ:self-attention-projection}
    \mathbf{Q_e} &= \mathbf{X}\mathbf{W^Q_e},\quad \mathbf{K_e} = \mathbf{X}\mathbf{W^K_e},\quad \mathbf{V_e} = \mathbf{X}\mathbf{W^V_e}
\end{align}
where $\mathbf{Q_e}, \mathbf{K_e}, \mathbf{V_e}$ are query, key, and value matrix after affine transformation by separately multiplying $\mathbf{X}$ with trainable parameters $\mathbf{W^Q_e}\in\mathcal{R}^{O\times D_{K}}, \mathbf{W^K_e}\in\mathcal{R}^{O\times D_{K}}, \mathbf{W^V_e}\in\mathcal{R}^{O\times D_{V}}$. 
\begin{align}\label{equ:self-attention-layernorm}
 \mathbf{H_e}  = \text{LayerNorm}(\mathbf{E_e} * \mathbf{W^O_e} + \mathbf{Q_e})
\end{align}
where $\mathbf{W^O_e}$ is the linear projection parameter. $\mathbf{H_e}\in\mathcal{R}^{|\mathbf{A}|\times D_K}$ is the encoding embedding matrix of input items $\mathbf{A} = \{a_i\}_{i=1}^N$. 

\textbf{Slate Proposer}
After the transformation and projection of the initial item list and context features, \textit{DivNet} will propose slate output by sequentially selecting items and arranging them in an optimal way which maximizes the utility and maintain diversity. The \textit{DivNet} achieves this goal by considering influences of previously-selected items on current selection and scoring of items as self-correcting process and enforce diversity during the learning and inference process.

Assuming selected item list $\big \{ I_{\pi_{1}},I_{\pi_{2}},\ldots, I_{\pi_{t-1}} \}$ up to time $t$, then the logit of the $t$th item candidate chosen from set $R_t:$ $\{ I_{\pi_{t}} \in \mathbf{A} : \{a_i\}_{i=1}^N, I_{\pi_{t}}\notin \{ I_{\pi_{1}},I_{\pi_{2}},\ldots, I_{\pi_{t-1}} \} \big \} $ can be computed as follows when considering the influence of selected items:

\begin{align}\label{equ:decode}
    \mathit{\hat{H}} &=  [H_{\pi_{1}},H_{\pi_{2}},\ldots,H_{\pi_{t-1}},H_{\pi_{t}}]\\
     \mathbf{Q_d} &= \mathit{\hat{H}}\mathbf{W^Q_d}, \quad \mathbf{K_d} = \mathit{\hat{H}} \mathbf{W^K_d}, \quad \mathbf{V_d} = \mathit{\hat{H}}\mathbf{W^V_d} \\
    \mathbf{\bar{H}} &= \text{Softmax}\big(\mathbf{Q_d} * \mathbf{K_d}^T * \mathbf{U}) \frac{\mathbf{V_d}}{\mathbf{\sqrt{D_K}}} \label{equ:embed}\\
    y_{\pi_{t}} &=  \sigma ( W_d \mathbf{\bar{H}}_{\pi_{t}} + b )
\end{align}
where $\mathbf{U}$ is upper triangular matrix which has all the elements below the main diagonal as zero and other elements as one. This ensure the causal influence where only current item depend on previously-selected items. $\sigma$ is sigmoid activation function. $y_{\pi_{t}}$ is the estimated utility logit for item $I_{\pi_{t}}$. To induce diversity in the recommendation list, self-correcting process is introduced in the following way. First, item similarity between item $i$ and $j$ is defined as:
\begin{align}
    L_{ij}=cosine(\mathbf{\bar{H}_i}, \mathbf{\bar{H}_j})
\end{align}
where $\mathbf{\bar{H}_i},\mathbf{\bar{H}_j}$ is the embedding of item $i$ and $j$ in Equation~\ref{equ:embed} and \textit{cosine} is the cosine distance between two vectors. Then the relevance of item $I_{\pi_t}$ to previously-selected items can be calculated using the following formula:
\begin{align}
    \det (I_{\pi_t}) = \begin{bmatrix}
    L_{\pi_{1}\pi_{1}} & L_{\pi_{1}\pi_{2}} & L_{\pi_{1}\pi_{3}} & \dots  & L_{\pi_{1}\pi_{t}} \\
    L_{\pi_{2}\pi_{1}} & L_{\pi_{2}\pi_{2}} & L_{\pi_{2}\pi_{3}} & \dots  & L_{\pi_{2}\pi_{t}} \\
    \vdots & \vdots & \vdots & \ddots & \vdots \\
    L_{\pi_{t}\pi_{1}} & x_{\pi_{t}\pi_{2}} & x_{\pi_{t}\pi_{3}} & \dots  & L_{\pi_{t}\pi_{t}}
\end{bmatrix}
\end{align}
$det(I_{\pi_t})$ is the determinant of matrice of L where 
element $L_{ij}=cosine(\mathbf{\bar{H}_i}, \mathbf{\bar{H}_j})$ is a
scaled measurement of the similarity between items i and j. $det(I_{\pi_t})$ increases when the item $I_{\pi_t}$ is less similar to previous items.
As the relevance of current item selection to previous items can be measured. The probability of selecting item $I_{\pi_t}$ at step $t$ can be computed:
\begin{align} \label{equ:item_prob}
    P(I_{\pi_t}) = \frac{y_{\pi_{t}} + \alpha det(I_{\pi_t})}{ \sum_{I_{\pi_t'} \in R_t} \big ( y_{\pi_t'} +  \alpha det(I_{\pi_t'}) \big) }
\end{align} 
The bigger utility of item $I_{\pi_t}$ and less similar to previous items, the higher probability the item will be selected. $\alpha$ is a parameter, governing the degree of diversity. For larger values, the DivNet tends to generate more diverse item list.

\subsection{Learning Algorithms}
To estimate the parameters of DivNet, one viable optimization is the widely-used reinforce algorithm~\cite{williams1992simple}. The objective is to maximize $\mathcal{R}(\pi, y)$ where $\pi$ is sampled from our \textit{DivNet} model and $y$ comes from logged click-through data:
\begin{align}\label{equ:reinforce}
    \max_\theta \mathbb{E}_{\pi\sim P_\theta}\big[\mathcal{R}(\pi, y)\big]
\end{align}
where $\theta$ denotes the parameters of \textit{DivNet}. The $\mathcal{R}(\pi, y)$ can be the evaluation metric measuring the users' satisfaction of item layout. In this paper, we use NDCG as $\mathcal{R}(\pi, y)$. Policy gradient is used to optimize the Equation~\ref{equ:reinforce} where the gradient is given by:
\begin{align}\label{equ:pg}
     \nabla_{\theta} \mathbb{E}_{\pi\sim P_\theta}\big[\mathcal{R}(\pi, y)\big] = \mathbb{E}_{\pi\sim P_\theta} \big[\mathcal{R}(\pi, y)\nabla \log P_{\theta}(\pi)\big] 
\end{align}
The Equation~\ref{equ:pg} can be approximated using Monte-Carlo method by sampling $\pi$ from $P_{\theta}$. 

As we already know that the sampling space is very large, optimizing from scratch using reinforce suffers from high variance and non-convergence. We indeed observe that the model occasionally converges to an abnormal point during training. To solve the out-of-distribution exposure bias problem,  we turn to supervised regularization with discounted objective which is robust and efficient. we can combine the loss of reinforce with this regularization.

Considering $t$st step of decoding process, DivNet assigns each item a score using Equation~\ref{equ:item_prob}.  We can define a per-step loss which uses multi-label cross-entropy loss.
\begin{align}
    \mathcal{L}_t = -\sum_{I_{\pi_{t'}}\in  R_t} y_{I_{\pi_{t'}}} log\textbf{P}(I_{\pi_{t'}})
\end{align}
where $I_{\pi_{t'}}\in  R_t$ means that at step t we only compute the multi-label cross-entropy loss on items which are not selected up to step t.

The final loss is the sum of per-step cross-entropy loss:
\begin{align}
    \mathcal{L} = \sum_{t=1}^{|A|} w_t \mathcal{L}_t
\end{align}
where we can optionally set weight $w_i=1/\log(t+1)$ for $t$-th step, like position discounting in NDCG. In summary, the final loss after addition of supervised regularization is as follows:
\begin{align}\label{equ:loss}
     \min_\theta - \mathbb{E}_{\pi\sim P_\theta}\big[\mathcal{R}(\pi, y)\big] +  \lambda\mathcal{L}
\end{align}
where $\lambda$ is the coefficient of the regularization term.
The overall learning procedure is summarized in Algorithm~\ref{alg:algorithm}.

\begin{algorithm}[tb]
\caption{Learning Algorithm of \textit{DivNet}}
\label{alg:algorithm}
\begin{algorithmic}[1] 
\STATE \textbf{Input}: training dataset $\mathcal{D}$
\STATE \textbf{Training Procedure}
\WHILE{Not Convergence}
\STATE Sample batch data {A,Y} from $\mathcal{D}$.
\STATE Feed samples into \textit{DivNet}, and the item list are sampled from DivNet using equation~\ref{equ:item_prob}. 
\STATE Update \textit{DivNet} using gradient descent by minimizing Equation~\ref{equ:loss}.
\ENDWHILE
\item[]
\STATE \textbf{Inference Procedure}
\STATE Given ranked items from upstream modules, pass them into \textit{DivNet} and output $\pi$ are sampled using Equation~\ref{equ:item_prob}.
\end{algorithmic}
\end{algorithm}

\section{Experiments}
In the section, we introduce the datasets used and experimental settings. Offline and online experimental results and discussions are also given. 

\subsection{Datasets}
Public datasets including Yahoo Webscope~\footnote{https://webscope.sandbox.yahoo.com} and Microsoft 10K~\cite{qin2013introducing} and real-world E-commerce dataset are used to testify models' performance. The summary of dataset statistics are given in Table~\ref{tab:data}.

\textbf{Yahoo Webscope dataset}
The dataset are collected from Yahoo search engine and made public to researchers by Yahoo for learning-to-rank method development. The total queries have the size of 29 thousands and the number of documents is 700 thousands. The document relevance scores given a query range from 0 (most irrelevant) to 4 (most relevant). The ratings (0-4) are converted into binary labels (0-2 negative and 3-4 positive) The documents contains textual features and statistics features. The queries are evenly sampled from the search engine to ensure the dataset is unbiased. 

\textbf{Microsoft 10K dataset}
The Microsoft 10K dataset is also collected from Bing search engine in a similar way to Yahoo dataset. It contains 10 thousands of queries and 1.2 millions of documents. 
The document also contain statistic features and relevance scores ranging from 0 to 4, denoting the matching score between queries and documents. The ratings (0-4) are also converted into binary labels (0-2 negative and 3-4 positive).

\textbf{E-commerce dataset} The large-scale dataset collected from a real-world E-commerce recommendation system contains records which have users, and recommended item list and user's response (click and non-click). Since different items have different color/icons and text descriptions, items would interact with surrounding items and influence users' responses.

\begin{table}
 \centering
 \renewcommand\arraystretch{1.1}
    \caption{Statistics of datasets used in this paper.} 
    \label{tab:data} 
    \begin{tabular}{c | c c  c}
      \hline
    \multirow{ 2}{*}{Datasets} &   \multicolumn{3}{c}{Data Attributes} \\
     \cline{2-4}
    &\#Users &\#Items &\#Samples \\
     \hline
    Yahoo   ~& & 29K  & 700K  \\
    \hline
    Microsoft ~& &  10K & 1.2M \\
    \hline
    E-commerce Data & 862K &  8.3M & 29.4M\\
    \hline
  \end{tabular}
\end{table}

\subsection{Experimental Setup}
\noindent\textbf{Experimental Settings}
For Yahoo and Microsoft dataset, the initial ranking lists are generated by LambdaMART. 
The dimensions $L$ of item embedding $\mathbf{I}\in \mathcal{R}^{|\mathbf{A}|\times M}$ and user embedding $\mathbf{U}\in \mathcal{R}^{|\mathbf{A}|\times K}$ used are both 64. The dimension $D_K, D_V$ of parameter matrix of key and value $\mathbf{W^K_e},\mathbf{W^V_e},\mathbf{W^K_d},\mathbf{W^V_d}$ are 64. Network parameters are initialized uniformly at random in [-0.1, 0.1].
 The stochastic gradient descent optimizer we used is ADAM~\cite{kingma2014adam} with an initial learning rate of 0.01. All experiments use batches of 256 training examples. The dataset are split into training, validation and testing by 0.8, 0.1 and 0.1. We test the parameter $\lambda$ from 0.01 to 1. 

\bigskip
\noindent\textbf{Baselines} For off-line evaluation, we compare the proposed method with ranking methods without collective recommendation and ranking methods with diverse collective recommendation.
\begin{itemize}
    \item \textbf{Wide\&Deep} This model represents recommendation system without collective recommendation module. The method used is a widely-used Wide\&Deep ranking model~\cite{cheng2016wide}. 
    
    \item \textbf{LambdaMART} This method~\cite{burges2010ranknet} is the SOTA learning-to-rank algorithm which uses listwise loss to model the contextual items when scoring items.
    
    \item \textbf{Submodular} This method~\cite{nassif2018diversifying} tries to balance the utility, such as click-through rate and item diversity by designing a sub-modular function. The items can be selected greedily by maximizing the sub-modular objective. 
    
    \item \textbf{DPP} This method~\cite{wilhelm2018practical} employs determinantal point processes to diversify recommendation set by computing pair-wise item similarity and iteratively select items from the learned determinantal point processes probabilistic model.
    
    \item \textbf{PRM}~\cite{pei2019personalized}  uses self-attention networks to encode contextual items and estimate their utilities based on contextual features. The items are reranked in a descending order based on the estimation of utility. 
      
    \item \textbf{Seq2Slate}~\cite{bello2018seq2slate}  uses recurrent neural networks to encode item interaction and sequentially output item ranking. Seq2Slate uses pointer network to sequentially generate the final list and the current item selection depends on previously-selected items. 

\end{itemize}

\bigskip
\noindent\textbf{Metrics}
To evaluate the performance of different methods, We use the user engagement metrics as the proxy for user overall satisfaction with the whole-page recommendation, which includes relevance and diversity.
Some of them are used in off-line evaluation and some of them are calculated by online A/B testing platform.  The definitions of those metrics are as follows:
\begin{itemize}
    
    \item \textbf{NDCG} is normalized discounted cumulative gain which measures the ranking quality. NDCG = DCG/IDCG, where DCG is calculated as the following equation:
    \begin{align}
        DCG = \sum_{i=1}^n \frac{2^{rel_i} - 1}{\log_2 (i+1)}
    \end{align}
    where $rel_i$ is the click label of $i$-th item. IDCG is ideal discounted cumulative gain and is equal to DCG after transforming $rel_i$ by sorting $rel_i$ decreasingly. 
    
    \item \textbf{Pre@K} is defined as the percentage of clicked items in the top-k following the user scan order. 
    \begin{align}
        Pre@K &= \frac{1}{\mathit{R}}\sum_{r\in \mathit{R}} \frac{\sum_i^K I(\mathit{S}_r (i))}{K}\\
        MAP@K &= \frac{1}{\mathit{R}}\sum_{r\in \mathit{R}} \frac{\sum_i^K \text{Pre@i} * I(\mathit{S}_r (i))}{K}
    \end{align}
    where $I(i)$ is the indicator function about whether the $i$-th item is clicked. R is the number of all user requests in the test dataset. $S_r$ is the ordered items generated by a ranking model.
    
\end{itemize}

\subsection{Offline Experiments}
In this section, we first conduct experiments on Yahoo, Microsoft and E-commerce datasets. Then ablation study is carried out to show the contribution of each part in \textit{DivNet}.
The results of different methods on three datasets are shown in Table~\ref{table:public}. 
\textit{LambdaMART} compares the utility scores of the list of items and performs better than \textit{Wide\&Deep} which neglects the influence of contextual items. Similarly, \textit{PRM} outperforms \textit{Wide\&Deep} by using the list of items as input features to model the impact of contextual items. The performances of \textit{Submodular} and \textit{DPP} vary in different datasets. \textit{DPP} works better than \textit{Submodular} in Yahoo and E-commerce datasets and worse than \textit{Submodular} in  Microsoft datasets, which shows the complexity of real-world dataset and 
the expertise knowledge in the model limits its expressive power.
\textit{Seq2Slate} and \textit{DivNet} outperforms alternatives with a large margin, which demonstrates that modeling user's sequential viewing behaviors is vital in deciding the item list. \textit{DivNet} improves NDCG by 3.6\% and MAP by 3.3\% over SOTA \textit{Seq2Slate} in E-commerce dataset and also improves in Yahoo and Microsoft datasets. 

To understand the effectiveness of the sub-modules of \textit{DivNet}, ablation test is carried out. The results are given in Table~\ref{table:ablation}. The performance of DivNet without Supervised learning (SL) where $\lambda$ equals zero is slightly worse than DivNet, which means that SL provides additional signal and guide the DivNet converge to superior region than DivNet without SL. Removing sampling by greedy decoding item when generating item lists deteriorates the model performance. The diversity of generated samples enlarges the data distribution of training samples and reduces the distribution gap between training and testing samples.

\begin{table}[t]
\centering
\caption{Offline performance on three datasets.} 
\begin{tabular}{c| c |  c c}
 \hline 
Dataset  & Algorithm& NDCG@10 & MAP@10  \\
  \hline

\cline{2-4}
        & Wide\&Deep & 0.7202 & 0.6954 \\
        & LamdaMART & 0.7346 & 0.7121   \\
        & Submodular  & 0.7274 & 0.7026   \\
 Yahoo	& DPP & 0.7379 & 0.7148   \\
        & PRM   & 0.7526 & 0.7223    \\
		& Seq2Slate   & 0.7562 & 0.7264    \\
		& DivNet &  \textbf{0.7619}  &  \textbf{0.7351}  \\		
 \hline
  
          & Wide\&Deep & 0.4261 & 0.4880  \\
          & LamdaMART & 0.4683 & 0.4971  \\
          & Submodular & 0.4562 & 0.4932  \\
Microsoft & DPP & 0.4536 & 0.4907  \\
          & PRM   & 0.4592 & 0.5143    \\
          & Seq2Slate & 0.4623 & 0.5214   \\
          & DivNet & \textbf{0.4756}  & \textbf{0.5298}  \\
  \hline

&Wide\&Deep  &0.3196  &0.2902\\
&LamdaMART  &0.3382  &0.3194 \\
&Submodular  &0.3367  &0.3151 \\

E-commerce &DPP       &0.3423  &0.3293 \\

 &PRM     &0.3435  &0.3304 \\ 

 &Seq2Slate     &0.3421  &0.3210 \\

 &DivNet     &\textbf{0.3562}  &\bf{0.3436}  \\
 \hline
\end{tabular}
\label{table:public} 
\end{table}

\begin{table}[h!]
\caption{Ablation test of different sub-modules of DivNet on E-commerce recommendation task.}
\begin{tabular}{|l |l |l|}
 \hline
 & \multicolumn{2}{c|}{Evaluation Metrics}\\
 \hline
Model & NDCG@10 & MAP@10\\
\hline
DivNet     &0.3562 &0.3436\\
\hline
Remove SL      &0.3493 &0.3402\\
\hline
Remove Sampling     &0.3514 &0.3416\\
\hline
\end{tabular}
\label{table:ablation}
\end{table}

\subsection{Online Deployment Results}
We also deploy the proposed $\textit{DivNet}$ for item recommendation in E-commerce recomendation platform.  The service is a marketing page that recommends items of various categories such as coupons and popular products to users. Our goal is to maximize the one-hop click-through rate by rearranging the whole-page item layout. We have a base ranker, a sophisticated and heavy DNN model who emphasizes on fine-ranking. After the base ranker, a simple rule checker is used to control duplicated items from the same category. We want to model mutual influence between items instead of simply taking category duplication into consideration via rules. Since different services have different color/icons and text descriptions, we believe different combinations will lead different results. In this way, we can provide users with items that they are interested in and improve the adhesiveness of our application. Due to the huge number of users, daily logged data is enormous. We use the business data collected over two months as our training and testing data. Some data statistics are given in Table~\ref{table:data_stat_mem}
\begin{table}[h!]
\caption{Statistics of sampled logged data from E-commerce platform for training and testing. UVs means the number of user-views and PVs represents the number of page-views.}
\begin{tabular}{|l |l |l|}
 \hline
 & \multicolumn{2}{c|}{Sampled Logged Data}\\
 \hline
 Types& Training & Testing\\
\hline
UVs  & 4.7M &  27k\\
 \hline
PVs  & 100M & 210k \\
 \hline
\end{tabular}
\label{table:data_stat_mem}
\end{table}

Results of online performances are shown in Table~\ref{table:calendar_online}. Compared to the baseline Wide\&Deep, $\textit{DivNet}$ has the highest page click-through rate improvement (+8.7\%) and user click-through rate (+4.26\%). We also observe other business indicators increase, such as user CVR. Now the \textit{DivNet} have been deployed in the E-commerce production task and is responsible for 100\% of user requests. 

\begin{table}[h!]
\caption{Online Performances of different approaches on E-commerce recommendation task.}
\begin{tabular}{|l |l |l|}
 \hline
 & \multicolumn{2}{c|}{Evaluation Metrics}\\
 \hline
Algorithm& page CTR lift & user CTR lift\\
\hline
LambdaMART     &+4.16\% &+3.05\% \\
\hline
SubModular     &+3.61\% &+2.69\%\\
 \hline
 DPP     &+6.7\% &+3.16\%\\
\hline
PRM     &+7.05\% &+3.60\%\\

 \hline
 Seq2Slate     &+7.4\% &+3.93\%\\
 \hline
 DivNet     &\bf{+8.7\%} &\bf{+4.26\%}\\
 \hline
\end{tabular}
\label{table:calendar_online}
\end{table}

\begin{figure}[htb]
\centering
\includegraphics[width=0.3\textwidth]{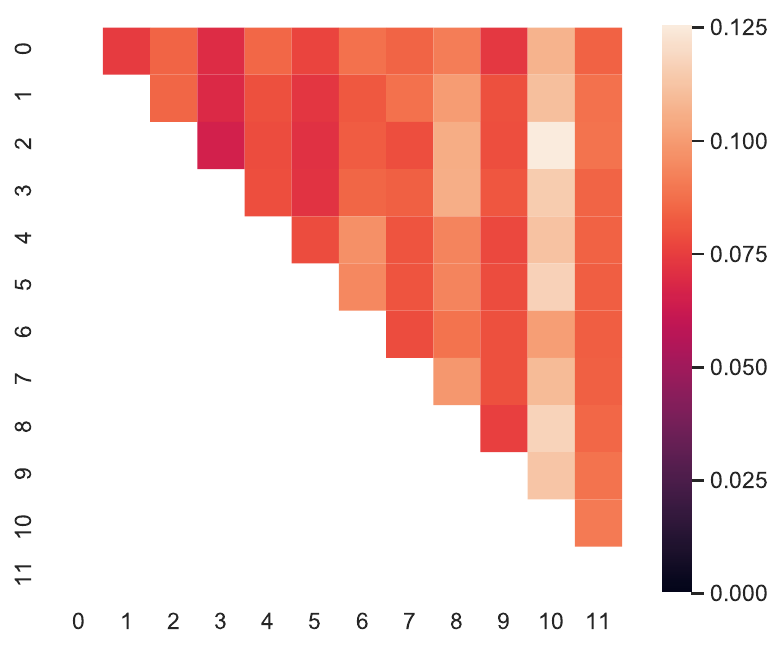}
\caption{Typical self-exciting pattern between items.}
\label{fig:influence}
\end{figure}

\noindent\textbf{Self-Correcting Effect}
One example of pair-wise influence in Equation $\frac{\text{Softmax}\big(\mathbf{Q_d} * \mathbf{K_d}^T*\mathbf{U})}{\mathbf{\sqrt{D_K}}}$ between items in E-commerce datset is visualized in Figure~\ref{fig:influence}. In this case, item 10 receive the most influence from previous items 2, 4, 8 where they share the same clothes category, which demonstrates items from the same category tend to have a larger self-correcting effect.

\section{Conclusion}
In this paper, we propose a computationally-efficient and effective item layout optimization model called $\textit{DivNet}$, which has been easily integrated into existing commercial recommendation systems. This model follows a sequence-to-sequence framework and explicitly consider the self-correcting effect when recommendation sets are sequentially viewed by users. The model can be learned end-to-end and is data-driven. Experimental results of off-line and online analyses show its superior performance. For the future work, incorporation of structure knowledge in \textit{DivNet} is a promising direction, e.g, making the objective function sub-modular. Multi-task objective optimization is also needed  as the business operators usually care about multiple business indicators. 

\bibliographystyle{ACM-Reference-Format}
\balance
\bibliography{ref}


\end{document}